\documentstyle[aps,eqsecnum,twocolumn]{revtex}
\draft
\begin{document}
\author{A. C. Doherty, S. M. Tan, A. S. Parkins, D. F. Walls}
\address{Department of Physics, University of Auckland, Private Bag
  92019, Auckland, New Zealand}
\title{State determination in continuous measurement}
\date{March 9, 1999}
\maketitle

\begin{abstract}
The possibility of determining the state of a quantum system after a
continuous measurement of position is discussed in the framework of quantum
trajectory theory. Initial lack of knowledge of the system and external
noises are accounted for by considering the evolution of conditioned density
matrices under a stochastic master equation. It is shown that after a finite
time the state of the system is a pure state and can be inferred from the
measurement record alone. The relation to emerging possibilities for the
continuous experimental observation of single quanta, as for example in
cavity quantum electrodynamics, is discussed.
\end{abstract}


\section{Introduction}

There is currently great interest in experiments which obtain useful
information about a quantum system or state in single runs of an experiment.
Very recently it has been possible to distinguish the quantum mechanical and
classical models of the interaction of an atom with a mode of a high finesse
optical cavity as the result of continuous monitoring of the output light
while a single atom passes through the cavity \cite{mabuchi1998b}. As a
result of this continuous monitoring the quantum mechanical backaction of
the measurement process may be expected to have a significant effect on the
evolution of individual runs of the experiment. Moreover, it may be possible
in the future to modify the evolution of the system through feedback based
on this continuous measurement \cite{wiseman1993c}. Current experimental
technology, such as that described in \cite{mabuchi1998b}, is reaching the
point at which determining the state of the system and observing the effects
of backaction and feedback in a single run of an experiment is a real
possibility. With this situation in mind, we discuss the identification of
the state of the system following a period of continuous observation and the
extent to which this state can be tracked, taking into account factors such
as imperfect initial knowledge of the state and imperfect detection
efficiency. We focus on a model of continuous position measurement of a
mechanical oscillator which is relevant to the experiment of Mabuchi {\em et
al.} but which also has relevance to other instances of interferometric
position monitoring such as gravitational wave detection.

The problem of describing quantum systems undergoing continuous measurement
has attracted much theoretical interest in recent years. As discussed by
Wiseman \cite{wiseman1996a} these theories admit a variety of
interpretations; as tools for efficient stochastic calculation of ensemble
averages in lieu of solving master equations \cite{molmer1993a}, as
equations describing the evolution of systems conditioned on measurements 
\cite{carmichael1993a,wiseman1993a,barchielli1993a} and as a description of
the evolution of a system coupled to an environment, in which collapse of
the wavefunction is supposed to be associated with the coupling to the
environment \cite{gisin1992a}. Here, we take the second viewpoint namely
that the conditioned state represents the observer's best description of the
system state given the results of the continuous measurement process.
Adopting the first or third viewpoints one is led to describe the system by
a pure state vector throughout the evolution although the reasons for doing
so are somewhat different in each case. By contrast, a description of one's
conditioned state of knowledge necessarily requires mixed states in order to
account for incomplete knowledge of the system. From this viewpoint the
fundamental equation for the conditioned evolution is the stochastic master
equation (SME) \cite{milburn1996a}. This is able to account for the effects
of mixed initial states, imperfect detection efficiencies and the existence
of unmeasured couplings to the environment. However, to date, relatively
little work has attempted to address the evolution of the conditioned state
in any of these situations \cite{mixsme}. In this paper we
consider a system which is simple enough that almost all the work can be
done analytically and which admits a treatment of all of these
imperfections. This helps in developing intuition about the role of SME's
and their possible relevance to experiments.

A projective measurement has the property that if the result of the
measurement is known, the state after the measurement is pure, and
depends 
only on the measurement result. It would be hoped that in a continuous
measurement there would be some finite interval of time after which the
measurement has effectively given rise to a projection, so that
the system is placed in a particular state which depends only on the
sequence of 
measurement results and which can be calculated without knowledge of the
initial 
state. If the resulting state is pure then a
stochastic Schr\"{o}dinger equation (SSE) would be a perfectly
adequate tool 
for describing the subsequent system evolution. In this paper we
investigate 
the conditions which lead to such an effective collapse and over what
timescale it takes place. This is made possible by considering density
matrices and the SME\ rather tha wave vectors and the SSE. In a real
experiment there will also be uncontrolled, unmeasured couplings of the
system to the environment and in this case the effects of the measurement
will compete not only with the coherent internal dynamics of the system but
also with the randomizing effects of the coupling to the bath. This may
lead to mixed conditioned states even after long periods of continuous
measurement and limit the observer's ability to make inferences about the
system state. Understanding the process by which the conditioned state
may collapse onto a pure state and the effects of noise as described
by the SME allows us to define conditions under which
continuous measurements in real experiments are approximations to
ideal measurements. 

This paper is organized as follows. In Sec.\ \ref{ssme} we establish our
simplified model of the continuous position measurement of an
oscillator and solve the SME for Gaussian
initial states. We find the time over which the second-order moments
approach
their steady-state values and calculate the entropy of the conditioned
state
as it becomes pure. In Sec.\ \ref{class} we discuss the classical problem of
state identification for the noisy measurement of the the position of an
oscillator and derive a kind of uncertainty principle relating the
observation and process noises if the classical model is to reproduce the
SME. In Sec.\ \ref{mrd} we show that the time-scale over which the
second-order moments of the conditioned state reach their steady state
is the
same as that over which the conditioned state is completely determined
by the measurement record.
Section~\ref{heat} discusses the effect of heating, noise and detection
inefficiency
on these conclusions. Finally in Sec.\ \ref{discuss} we summarize and
make
some comments about future extensions of this work.

\section{Solving the Stochastic Master Equation} \label{ssme}

\subsection{A Generalized Model for Continous Position Measurement}

In this paper we shall consider the abstract model of continuous position
measurement discussed by Caves and Milburn \cite{caves1987a,milburn1996a}
with a harmonically bound rather than a free measured
particle. Projective
position measurements are imagined to be made on a sequence of meters
coupled briefly to the system and the limit of very frequent meter
interactions and very
broad initial
meter position distributions is taken. This leads to a continuous
evolution
for the system of interest. Although this model should correspond in some
limit to any continuous position measurement of a single oscillator at
the
standard quantum limit, one system which does realize it at least
approximately is the dispersive regime of single atom cavity quantum
electrodynamics (QED). In this system the position of an atom inside a
high-finesse optical cavity causes a phase shift of the field driving the
cavity which can be monitored by homodyne detection 
\cite{storey1992a,milburn1994a,quadt1995a,doherty1998a}. The SME for the
conditioned evolution of the system in its It\^{o} form is 
\cite{wiseman1994a,milburn1996a} 
\begin{equation}
d\rho _{\text{c}}=-i[H_{\text{sys}},\rho _{\text{c}}]dt+2\alpha {\cal D}%
[x]\rho _{\text{c}}dt+\sqrt{2\alpha }{\cal H}\left[ e^{-i\phi }x\right]
\rho_{\text{c}}dW. \label{SME1}
\end{equation}
The superoperators ${\cal D}[c]$ and ${\cal H}[c]$ acting on a density
matrix $\rho $ are ${\cal D}[c]\rho =c\rho c^{\dagger }-\frac{1}{2}%
c^{\dagger }c\rho -\frac{1}{2}\rho c^{\dagger }c$ and ${\cal H}[c]\rho
=c\rho +\rho c^{\dagger }-$Tr$(c\rho +\rho c^{\dagger })\rho $ where
$c$ is
an arbitary operator. We will imagine that the atom is harmonically
bound,
\begin{equation}
H_{\text{sys}}=\frac{p^{2}}{2m}+\frac{m\omega ^{2}x^{2}}{2}.
\end{equation}
The constant $\alpha $ describes the strength of the measurement interaction
and in the cavity QED example depends on the strength of the coherent
driving and of the damping of the cavity. For the moment we will
consider a one-sided cavity and perfect detection so that all the
output light is detected. This assumption will be relaxed in Sec.\
(\ref{heat}). In a generalization of the
Caves and Milburn model we
will allow projective measurements of any quadrature of the meters, not just
position, since this can be realized in the cavity QED experiments by
varying the phase $\phi $ of the local oscillator in the homodyne detection
of the output light. The resulting measurement current $i=\frac{dQ}{dt}$
(suitably scaled)\ is 
\begin{equation}
dQ=\cos (\phi )\langle x\rangle _{\text{c}}dt+\sqrt{\frac{1}{8\alpha }}dW.
\end{equation}
This stochastic master equation with the full dependence on
$\phi$ was discussed by Diosi \cite{diosi1988a} in the context of a phenomenological
model of position measurement through photon scattering where the kind
of measurement made on the scattered photon determines the value of
$\phi$. Clearly if we choose $\phi =0,\pi $ the homodyne detection is an effective
measurement of the atomic position, whereas if $\phi =\pi /2,3\pi /2$ the
measurement results are independent of the system state and only contain
information about the noisy potential seen by the atom. For $\phi =\pi
/2,3\pi /2$ the conditioned evolution is linear in the system state \cite
{milburn1996a,doherty1998a}. This is somewhat like a continuous quantum
eraser where the continuous measurement of one quadrature of the probe
system destroys the position information written into the other quadrature
of the probe. For $\phi =0$ the SSE corresponding to this model in the case
of pure states of the system has been considered in \cite
{belavkin1989a,jacobs1998a}. In this work we will use a simpler means of
solving the evolution equations which straightforwardly applies to mixed
states.

It has been shown by Jacobs and Knight \cite{jacobs1998a} that the SSE
corresponding to Eq.~(\ref{SME1}) is one for which Gaussian pure states \cite
{schumaker1986a} remain Gaussian and pure under the evolution. Thus,
if the system is
initially in a mixture of Gaussian pure states, the conditioned state will
remain Gaussian under the SME (\ref{SME1}). This property holds true for
single mode systems where the Hamiltonian is at most quadratic in $x$ and $p$
and the operator $c$ appearing in the Lindblad term ${\cal D}[c]$ is
linear in $x$ and $p$ 
and in all likelihood for multimode linear systems also. If we restrict
ourselves to Gaussian initial states --- for example to thermal states of
the oscillator --- then there are only five quantities which completely
define the state --- $\langle x\rangle $,$\langle p\rangle $,$\langle
(\Delta x)^{2}\rangle =\langle x^{2}\rangle -\langle x\rangle ^{2}$,$\langle
(\Delta p)^{2}\rangle =\langle p^{2}\rangle -\langle p\rangle ^{2}$ and $%
\langle \Delta x\Delta p\rangle _{\text{sym}}=\frac{1}{2}(\langle
xp+px\rangle -2\langle x\rangle \langle p\rangle )$. From now on we will use 
$\langle c\rangle $ to indicate the conditioned expectation value Tr$(c\rho
_{\text{c}})$. The requirement that the initial states be Gaussian is not
unduly restrictive since these are typically the states that are most stable
in linear systems and would therefore be a reasonable assumption for an
initial state. Moreover, there is numerical evidence that non-Gaussian pure
initial states become approximately Gaussian under the stochastic
Schr\"{o}dinger equation evolution corresponding to the SME (\ref{SME1}) on
timescales fast compared to those considered here \cite
{rigo1997a,garraway1994b}. It has also been shown that in at least one
such linear system an
arbitary density matrix can eventually be written as a probabilistic
sum over Gaussian pure states after sufficient evolution subject to
the unconditioned master equation \cite{halliwell1995a}. In considering such a linear system we are in
effect specifying a semiclassical evolution, since the equations of motion
for the Wigner function of the state in the unconditioned evolution are
similar to the classical Liouville equations for a phase space
density, see 
for example \cite{peres1993a}. However we here include the important
quantum feature of the 
measurement backaction, which is represented by the fact that the
momentum  
diffusion is determined by the accuracy with which the particle's
position 
is monitored \cite{milburn1996a}. So, as the noise in the position
measurement is decreased by increasing $\alpha ,$ the momentum diffusion
expressed by the Lindblad term in the SME (\ref{SME1}) increases.

Following Breslin and Milburn \cite{breslin1997a} we can derive a
system of 
differential equations for the first and second-order moments from the
SME (%
\ref{SME1}) since $d\langle c\rangle =$Tr$(cd\rho _{\text{c}})$. A
similar calculation is performed in \cite{halliwell1995a}. We define
the dimensionless quantities $\tilde{x} =$ $\langle
x\rangle/\sqrt{\hbar/2m\omega},\tilde{p}
=\langle p\rangle/\sqrt{\hbar m\omega/2 }$, the second-order moments
$V_{xx} =2m\omega \langle (\Delta x)^{2}\rangle/\hbar$, $V_{pp} =2\langle
(\Delta p)^{2}\rangle/\hbar m\omega$, $V_{xp}=2\langle \Delta x\Delta
p\rangle _{\text{sym}}/\hbar$ and a dimensionless parameter describing
the relative strengths 
of the measurement and harmonic dynamics, $r=m\omega ^{2}/2\hbar
\alpha$. The Heisenberg uncertainty principle now requires that $%
V_{xx}V_{pp}\geq 1$. A pure state has the property that $%
V_{xx}V_{pp}-V_{xp}^{2}=1,$ representing the fact that a Gaussian pure state
is a minimum uncertaintly state for some pair of conjugate quadrature
variables \cite{schumaker1986a}. In terms of these new quantities, the
It\^{o} stochastic differential equations for the first and second-order
moments and the measured photocurrent are 
\begin{mathletters}
\label{momev} 
\begin{eqnarray}
d\tilde{x} &=&\omega \tilde{p}dt+\sqrt{\frac{2\omega }{r}}\cos (\phi
)V_{xx}dW,  \\
d\tilde{p} &=&-\omega \tilde{x}dt+\sqrt{\frac{2\omega }{r}}\left( \cos (\phi
)V_{xp}-\sin (\phi )\right) dW, \\
d\tilde{Q} &=&\cos \left( \phi \right) \tilde{x}dt+\sqrt{\frac{r}{2\omega }}%
dW, \\
\frac{1}{\omega }\frac{dV_{xx}}{dt} &=&2V_{xp}-\frac{2}{r}\cos ^{2}(\phi
)V_{xx}^{2}, \\
\frac{1}{\omega }\frac{dV_{pp}}{dt} &=&-2\left( 1-\frac{\sin (2\phi )}{r}%
\right) V_{xp}+\frac{2}{r}\cos ^{2}(\phi ) \nonumber \\
&&-\frac{2}{r}\cos ^{2}(\phi )V_{xp}^{2}, \\
\frac{1}{\omega }\frac{dV_{xp}}{dt} &=&V_{pp}-\left( 1-\frac{\sin (2\phi )}{r%
}\right) V_{xx} \nonumber \\
&&-\frac{2}{r}\cos ^{2}(\phi )V_{xx}V_{xp}.
\end{eqnarray}
\end{mathletters}
As in \cite{breslin1997a}, the It\^{o} rules for stochastic differential
equations \cite{gardiner1985a} and the properties of Gaussian states \cite
{gardiner1991a} result in deterministic equations for the conditioned
second-order moments which are decoupled from the equations for the
means. The
constant term in the equation for $V_{pp}$ refers to the momentum diffusion
due to the position measurement and remains in the master equation for
the 
unconditioned evolution. The non-linear terms describe the
conditioning of 
the state on the measurement. The noisy contribution to the equation
for $d%
\tilde{x}$ seems a little like a stochastic impulsive force, however
it is 
perhaps better to think of this term as updating the expected position
given 
the measurement result $d\tilde{Q}$ in analogy with classical Bayesian
state    
estimation.

Equations like those above for the second-order moments of the
conditioned 
state arise very frequently in classical continuous-time observation and
control problems. They can be collected into a Riccati matrix differential
equation \cite{reid1972a} for the covariance matrix,
\begin{mathletters} 
\begin{eqnarray}
\frac{d}{dt}V &=&\omega \left( C-VBV-DV-VA\right) ,  \label{riccati} \\
V &=&\left( 
\begin{array}{ll}
V_{xx} & V_{xp} \\ 
V_{xp} & V_{pp}
\end{array}
\right) ,  \\
A &=&D^{T}=\left( 
\begin{array}{ll}
0 & \left( 1-\frac{\sin (2\phi )}{r}\right) \\ 
-\left( 1-\frac{\sin (2\phi )}{r}\right) & 0
\end{array}
\right) , \\
B &=&\left( 
\begin{array}{ll}
\frac{2\cos \phi }{r} & 0 \\ 
0 & 0
\end{array}
\right) , \\
C &=&\left( 
\begin{array}{ll}
0 & 0 \\ 
0 & \frac{2\cos \phi }{r}
\end{array}
\right) .
\end{eqnarray}
\end{mathletters}
A single variable Riccati equation which arose from the
stochastic Schr\"{o}dinger equation for this system was found
and solved in \cite{belavkin1989a}.

In practice it may not be homodyne but rather heterodyne detection which can
be experimentally achieved with noise at the quantum limit \cite
{mabuchi1998b}. In this case, the local oscillator is detuned from the
cavity frequency by a frequency $\Delta _{\text{het}}$ which is large
compared to all system frequencies, with the result that the phase $\phi $
changes very rapidly. The quantum theory of heterodyne detection is
described by Wiseman and Milburn \cite{wiseman1993b}. The appropriate
conditioned evolution can be described by averaging all the trigonometric
functions of $\phi $ in the evolution equations (\ref{momev}) except where
they are multiplied by It\^{o} increments. Thus the equations for the
second 
order moments are exactly those for homodyne detection with $\phi =0$
where $%
r$ is replaced by $2r,$ corresponding to halving the signal to noise
ratio 
of the measurement. Considering the stochastic integrals
$\int_{t}^{t+\delta 
t}\cos (\Delta _{\text{het}}t^{\prime })dW(t^{\prime
}),\int_{t}^{t+\delta 
t}\sin (\Delta _{\text{het}}t^{\prime })dW(t^{\prime }),$ in the limit
firstly of infinite $\Delta _{\text{het}}$ then of infinitesimal
$\delta t,$ 
leads to equations for the first order moments under heterodyne detection 
\begin{mathletters}
\begin{eqnarray}
d\tilde{x} &=&\omega \tilde{p}dt+\sqrt{\frac{\omega }{r}}V_{xx}dW_{1}, \\
d\tilde{p} &=&-\omega \tilde{x}dt+\sqrt{\frac{\omega }{r}}\left(
V_{xp}dW_{1}-dW_{2}\right) ,
\end{eqnarray}
\end{mathletters}
where $dW_{1}$ and $dW_{2}$ are independent Wiener increments. Again
this is 
formally identical to the evolution with homodyne detection and $\phi =0$
where $r$ is replaced by $2r$ and in which a second independent noise
process $dW_{2}$ is present. Scaled versions of the two quadrature
components of the experimental photocurrent are given by 
\begin{mathletters}
\begin{eqnarray}
d\tilde{Q}_{1} &=&\tilde{x}dt+\sqrt{\frac{r}{\omega }}dW_{1}, \\
d\tilde{Q}_{2} &=&dW_{2}.
\end{eqnarray}
\end{mathletters}
Again note the replacement of $r$ by $2r$ in the equation for $d\tilde{Q}%
_{1} $ as compared to the equation for $d\tilde{Q}$ with $\phi =0$. So if
the quadrature-phase $I_{2}(t)$ current is collected and used to
account for the noisy 
potential, or alternatively fed back in order to compensate this
evolution, 
then heterodyne detection is equivalent to homodyne detection with
half the 
signal to noise ratio as far as the motional state is concerned.

\subsection{Steady State Conditioned Variances}

For all phases of the local oscillator $\phi \neq \pi /2,3\pi /2$ the
second-order moments possess a steady state. For example if $\phi =0$%
\begin{mathletters}
\begin{eqnarray}
V_{xx}^{\text{ss}} &=&\frac{1}{\sqrt{2}}r\sqrt{\sqrt{1+\frac{4}{r^{2}}}-1},
\\
V_{pp}^{\text{ss}} &=&\frac{1}{\sqrt{2}}r\sqrt{1+\frac{4}{r^{2}}}\sqrt{\sqrt{%
1+\frac{4}{r^{2}}}-1}, \\
V_{xp}^{\text{ss}} &=&\frac{1}{2}r(\sqrt{1+\frac{4}{r^{2}}}-1),
\end{eqnarray}
\end{mathletters}
which defines a pure state that agrees with the solution given in \cite
{belavkin1989a}. The steady states are found to have exactly the same
second 
order moments regardless of the initial purity of the system. Assuming
ideal 
detection, the observer is eventually able to ascribe a pure state to the
system. When the harmonic oscillator dynamics dominates over the
measurement 
($r\gg 1$)$\,$the steady conditioned state is approximately a coherent
state 
with $V_{xx}^{\text{ss}}\simeq V_{pp}^{\text{ss}}\simeq
1,V_{xp}^{\text{ss}%
}\simeq 1/r.$ If the measurement dynamics dominate ($r\ll 1$) then
$V_{xx}^{%
\text{ss}}\simeq \sqrt{r},V_{pp}^{\text{ss}}\simeq \frac{2}{\sqrt{r}}%
,V_{xp}^{\text{ss}}\simeq 1\,$and the conditioned state is strongly
squeezed 
in position, as one would expect for a position measurement which is
rapid 
enough to overcome the internal dynamics of the system. The product of
the position and momentum variances is greater than that required by
the Heisenberg uncertainty principle as a result of the internal
Hamiltonian which gives a non-zero correlation $V_{xp}$. 
Finally, the scaling we have chosen for the variances makes the limit
of the 
free particle appear singular, however this limit exists and the results
agree with Belavkin's \cite{belavkin1989a}. Pure steady states also exist
for other values of $\phi $ but the full expressions are rather
complicated 
so we will just consider two special cases. If the oscillator dynamics
dominate 
($r\gg 1 $) the steady conditioned states are insensitive to the local
oscillator 
phase leaving the conditioned state nearly in a coherent state $V_{xx}^{%
\text{ss}}\simeq 1+\sin (2\phi )/2r,V_{pp}^{\text{ss}}\simeq 1-\sin (2\phi
)/2r,V_{xp}^{\text{ss}}\simeq \cos ^{2}\phi /r.$ If $r\ll 1$ the
conditioned state is strongly dependent on the choice of local
oscillator phase, 
\begin{mathletters}
\begin{eqnarray}
V_{xx}^{\text{ss}} &\simeq &\frac{\sqrt{r}}{\left| \cos \phi \right| }\sqrt{%
1/\left| \cos \phi \right| +\tan \phi }, \\
V_{pp}^{\text{ss}} &\simeq &\frac{2}{\sqrt{r}}\sqrt{1/\left| \cos \phi
\right| +\tan \phi }, \\
V_{xp}^{\text{ss}} &\simeq &1/\left| \cos \phi \right| +\tan \phi .
\end{eqnarray}
\end{mathletters}
Rather surprisingly it is possible for some phases of the local
oscillator that the 
momentum variance is in fact smaller than the position variance. This is
because the measurement for non-trivial $\phi $ is a simultaneous
measurement of the position of the oscillator and the momentum kicks to
which it is being subjected and this can result in a more sharply defined
momentum than position. This is really only a possibility for phases of the
local oscillator where there is very little position information in the
record.

In the cases $\phi =\pi /2,3\pi /2$ the differential equations for the second
order moments of the state are simply those that result from the unitary
evolution of a simple harmonic oscillator and the SME (\ref{SME1}) describes
a stochastic unitary evolution. As a result there is no steady state for the
moments, which grow according to the conventional Schr\"{o}dinger equation.
In this case the measurement current is white noise. In \cite{rigo1997a} it
was noted that where a Lindblad operator is hermitian there exists an
unravelling of the master equation which does not localize the conditioned
state. The reason for this is clear in this context, such an unravelling
corresponds to a measurement in which the observer obtains no information
about the system state.

\subsection{Timescale for Determination of a Pure State through Measurement}

The matrix Riccati equation (\ref{riccati}) has an analytic solution given
by Reid \cite{reid1972a}. Where $U(t)$ and $W(t)$ obey the linear coupled
matrix equations
\begin{mathletters} 
\begin{eqnarray}
\frac{d}{dt}U &=&AU+BW, \\
\frac{d}{dt}W &=&CU-DW
\end{eqnarray}
\end{mathletters}
and for times where $U(t)$ is non-singular, the solution for the covariance
matrix is $V(t)=W(t)U^{-1}(t).$ The full solution for all the second-order
moments and arbitary initial conditions is complicated and not particularly
illuminating. In order to expose the general form of the solution we will
just consider the position variance in the case of an initial state of the
oscillator with $V_{xx}(0)=V_{pp}(0)=V_{0},V_{xp}(0)=0,$
\begin{mathletters}
\begin{eqnarray}
V_{xx}(t) &=&\frac{%
\begin{array}{c}
V_{0}\left( c^{2}\cosh 2bt+b^{2}\cos 2ct\right)  \\ 
+ \left( 1+V_{0}^{2}\right) \left( c\sinh 2bt-b\sin 2ct\right)
\end{array}
}{%
\begin{array}{c}
\left( b^{2}+c^{2}\right) +V_{0}\left( c^{3}\sinh 2bt+b^{3}\sin 2ct\right)
\\ 
+\left( 1+V_{0}^{2}\right) \left( c^{2}\sinh ^{2}bt-b^{2}\sin ^{2}ct\right)
\end{array}
}, \\
b &=&\frac{1}{\sqrt{2}}\sqrt{\sqrt{1+\frac{4}{r^{2}}}-1}, \\
c &=&\frac{1}{\sqrt{2}}\sqrt{\sqrt{1+\frac{4}{r^{2}}}+1}=1/V_{xx}^{\text{ss}%
}.
\end{eqnarray}
\end{mathletters}
We have scaled time by the harmonic oscillation frequency $\omega
$. When $2bt\gg 1$ the system is close to its steady state value
regardless of $V_{0}$. 
The non-linearity of the terms describing the conditioning of the system
state cause the time over which the conditioned state becomes pure to be
independent of the initial temperature of the state. For definiteness we
will define a collapse time $\tau _{\text{col}}=2/b\omega $ as being the
time at which the state has become effectively pure. When $r>1$ this
collapse time is $\tau _{\text{col}}\simeq 2r/\omega =m\omega 
/\hbar \alpha $ since in this regime $b\simeq 1/r$. In this regime
$c\simeq 
1+1/2r^{2}$ and there are many oscillations of the particle before the
state 
is determined. That the time to determine a particular pure state of the
system should increase with the frequency seems reasonable since
unexpected 
values of the measurement current could be due to a mistaken idea of the
position of the particle, the white noise in the measurement record or to
motion due to the oscillation and these possibilities will be more
difficult 
to distinguish if the atomic motion is fast. For smaller $r$, the
measurement is becoming very good and this estimate for the collapse
time is 
optimistic since it is hard to determine the state of the system
in less 
than one period of the mechanical oscillation. Reductions in the
conditioned 
momentum variance will only occur as the Hamiltonian evolution creates a
correlation between the position and the momentum of the state. Even
if the harmonic potential were absent a continuous measurement of
position will give some
information about the momentum of the particle. When $r\ll 1$ the
particle 
is essentially free as far as the measurement is concerned and the
time for 
the state reduction to occur turns out to be $\tau _{\text{col}}\simeq \sqrt{%
8m/\hbar \alpha }$ which is determined solely by the strength of the
measurement and the mass of the particle. In this situation $b\simeq c\simeq
1/\sqrt{r}$. Increasing the measurement coupling means that the time for the
measurement to take place is reduced and in the limit of infinite $\alpha $
the model essentially describes a projective measurement of the position. If
heterodyne detection is used rather than homodyne detection then $r$ is
replaced by $2r$ in the above equations with obvious implications for the
timescale of the system collapse.

A pure state describes a situation in which an observer has maximal
information about the system. A mixed state describes less than maximal
knowledge of system and it turns out that the amount of missing
information, in an information theoretic sense, required to complete
the specification of the state may be measured by the von Neuman
entropy $S(\rho )=-$Tr$(\rho \log \rho )$ of the density operator,
see for example \cite{caves1996a}. Thus the entropy allows us to quantify
the extent to which the measurement has determined the state of the system
at a given time and also the extent to which other environmental couplings
limit what an experimenter can say about the system state. Another commonly
used measure of the ``mixedness'' of a given density matrix is the linear
entropy or purity $s(\rho )=1-$Tr$(\rho ^{2})$. For a single mode Gaussian
state these quantities are simple functions of the unitless ``area'' of the
state in phase space $A=\sqrt{V_{xx}V_{pp}-V_{xp}^{2}}$ \cite{zurek1993a}
\begin{mathletters} 
\begin{eqnarray}
s(\rho ) &=&1-\frac{1}{A}, \\
S(\rho ) &=&\frac{A+1}{2}\ln (A+1)-\frac{A-1}{2}\ln (A-1)-\ln 2.
\end{eqnarray}
\end{mathletters}
Note that $A$ is just the determinant of the covariance matrix of the
position and momentum probability distributions for the conditioned
state. For a pure state $A=1,s(\rho )\rightarrow 0,S(\rho )\rightarrow
0$ and as 
the state becomes increasingly mixed it occupies a larger phase space
area 
such that as $A\rightarrow \infty ,s(\rho )\rightarrow 1,S(\rho
)\rightarrow 
\infty $. As we would expect if the state is widely spread in phase space
then our knowledge of the system is poor and the information needed to
complete the description of the state is large.

The time evolution of the variances and the linear and von Neumann entropies
of the conditioned state with $V_{xx}(0)=V_{pp}(0)=20,V_{xp}=0,\omega
=1,r=20 $ is plotted in Fig.\ (\ref{fig1}). These parameters are chosen
since the measurement dynamics are not fast enough to obscure the
harmonic oscillation totally and because achieving very small values
of $r$ will
probably be difficult in practice. Several features of Fig.\ (\ref{fig1})
are relevant. Firstly the initial very rapid reduction of the position
variance is associated with the the first part of the measurement record
making a reasonably accurate determination of the position. Then over the
timescale of the harmonic oscillation the momentum variance also
reduces as 
the dynamics correlate the position and momentum. Note that the
reduction in 
momentum variance occurs only when there is a strong correlation
between the 
position and the momentum. As $V_{xp}$ becomes small the position
variance 
decays more rapidly and the reduction of the momentum variance slows.
Eventually all the second-order moments decay to the steady state values
predicted 
above. This initial fast reduction of the position variance is
accompanied 
by a fast reduction of the von Neumann entropy which damps to zero as the
system approaches steady state.

\subsection{Cavity QED Realization}

Although we have been considering an abstract model for continuous
position 
measurement this work is motivated by emerging experimental
possibilities in 
areas such as cavity quantum electrodynamics (QED). The position
dependent 
phase shift induced by an atom strongly coupled to a high finesse optical
cavity mode in the dispersive limit of cavity QED \cite
{storey1992a,milburn1994a,quadt1995a,doherty1998a} realizes the abstract
position measurement coupling considered here. It is currently
possible to 
detect the presence of a single cold atom in the cavity through
measurements of the output field \cite{mabuchi1996a,hood1998a} and great
progress has been made in observing single atom events with broad
bandwidth 
close to the dispersive regime \cite{mabuchi1998b}. The phase shift
changes 
most quickly with position where the gradient of the field is greatest
and 
if the atom is harmonically confined in this region then the model
discussed 
above would be approximately realized. There is also the far off-resonant
optical potential which will lead to large forces on the atom in this
regime 
which would move the atom quickly away from this region of the standing
wave. However an optical standing wave at a nearby frequency could in
principle be tuned to cancel the ac-Stark shift of the ground state in the
region of the harmonic confinement. In fact the dipole force could
straightforwardly be included in our simplified model resulting in only
minor changes for sufficiently strong harmonic confinement. For the
moment 
we will not specify the exact source of the potential confining the
atom, but 
the use of light forces from a far off-resonant optical standing wave
or a 
standing wave in another mode of the cavity are possibilities. There
is also 
experimental work aimed at confining ions in high finesse optical
cavities 
which could realize such a system \cite{ions}. 

If we
imagine a harmonic potential confining the atom to this region of the
standing wave with some constant restoring force to overcome the dipole
force then in the far-detuned and Lamb-Dicke limit the resulting SME
would 
be exactly Eq. (\ref{SME1}) above \cite{doherty1998a}. The constant
$\alpha $ 
would then be equal to $2g_{0}^{4}nk_{L}^{2}/\Delta ^{2}\kappa $,
where $%
g_{0}$ is the maximal single photon Rabi coupling in the cavity, $n$
is the 
number of photons present in the driven cavity in the steady state,
$\Delta $ 
is the detuning between the atomic and cavity resonances --- the external
laser driving is on the cavity resonance, $\kappa $ is the cavity field
decay rate and $k_{L}\,$is the wave number for the light resonant
inside the 
cavity. Using the parameters of \cite{doherty1998a} which are based on
cavity parameters achieved by Hood {\em et al.} \cite{hood1998a} gives $%
\alpha =2.4\times 10^{20}$~s$^{-1}$m$^{-2}$ and this determines the
rate of 
decay of the off-diagonal terms in the position representation of the
density matrix under the unconditioned evolution. This rather large
number 
means that the density matrix elements $\langle x|\rho |x^{\prime
  }\rangle 
\, $where $x$ and $x^{\prime }$ are separated by nine nanometers or
around 
1\% of a wavelength will decay in the unconditioned evolution at the
rate $2.0\times 10^{4}\,$Hz. The decay of off-diagonal elements of the
density 
matrix in a particular basis is often associated with decoherence and the
emergence of classicality \cite{zurek1981a}. In this case the
decoherence is due in to the
measurement coupling. In \cite{milburn1994a} this decay of off-diagonal
density matrix elements is described as state reduction, in this work
we are 
interested in state reduction onto a pure state and the rate of this
process 
is determined not only by $\alpha $ but also by the length scale of a
typical pure state of the uncoupled system. Thus we found above that
reduction onto a pure state took place in a time $\tau
_{\text{col}}\simeq 
m\omega /4\hbar \alpha $ and the dependence on the length scale of the
harmonic oscillator is clear. In order to find the rate of collapse
onto a 
pure state in the conditioned evolution it is therefore also necessary to
know the oscillation frequency of the atom due to its harmonic
confinement. 
Assuming this is achieved optically, the value for $\omega /2\pi $
could be 
in the range of tens to hundreds of kilohertz. So for example in \cite
{doherty1998a} the potential for a cesium atom resulting from the same
cavity and driving parameters gives $\omega /2\pi =2\sqrt{\hbar
k_{L}^{2}/2m}\sqrt{g_{0}^{2}n/\Delta }/2\pi =180$~kHz, while $%
\omega /2\pi \simeq 60$ kHz has been achieved for cesium in optical lattices
by Haycock {\em et al. }\cite{haycock1997a}. For such a hypothetical
experiment with cesium we now have $r=5.6$ and $r=0.63$ respectively.
Estimates, as outlined above, for the time for an experimenter to determine
a pure state of the system through heterodyne detection are then $19\mu $s
and $8.9\mu $s. Both these times are reasonably close to the minimum
collapse time for this accuracy of detection which corresponds to the free
particle limit with $\tau _{\text{col}}\simeq 3.9\mu s.$ In current
experiments detection efficiencies and bandwidths will have a significant
effect on the information that can be gathered from the record. For trapped
ions harmonic frequencies would be around an order of magnitude larger and
state reduction times would then also be around an order of magnitude
longer. However the cavity finesse used here for such experiments may be
more difficult to achieve at typical frequencies of ion transitions
and the size of the cavity will be limited by the ion trap electrodes. What is dramatic about this time is that it is so
short, single cold atoms have been observed close to the centre of the
cavity mode in the experiment at Caltech for times of the order of hundreds
of microseconds. This is another confirmation of the extent to which this
experiment operates near the Standard Quantum Limit of position measurement (%
\cite{mabuchi1998a} and references therein) and an indication of the
importance of continuous quantum measurement theory to its interpretation.

Note that it will be difficult to achieve very small values of $r$ for
which 
the measurement dynamics dominate over the oscillation in the
well. This is 
the result of the two main constraints on the applicability of the model.
Firstly it is necessary that the harmonic potential confine the atom
to well 
within a wavelength, thus justifying the Lamb-Dicke approximation in the
master equation (\ref{SME1}). this requires that the recoil frequency for
the cavity transition $\omega _{\text{rec}}=\hbar k_{L}^{2}/2m$ be much
smaller than the harmonic oscillation. Moreover attaining the dispersive
regime requires that the saturation parameter $s=g_{0}^{2}n/\Delta
^{2}$ be 
much smaller than one. It is possible to express $r$ in terms of these
quantities: $r=(\omega /\omega _{\text{rec}})(1/8s)(\omega /\Gamma
_{\text{%
cav}})\,$where $\Gamma _{\text{cav}}=g_{0}^{2}/\kappa $ is the cavity
mediated spontaneous emission rate \cite{cirac1992b}. Assuming that values
of $\omega \,$and $s$ are chosen to satisfy a particular level of
approximation, the rate at which the conditioned states become pure is only
increased by changing the cavity parameters through increasing $\Gamma _{%
\text{cav}}$ relative to the oscillation frequency. Moreover, since both the
first two factors in this expression for $r$ must be large for the SME (\ref
{SME1}) to correspond to the cavity system, extremely large values of $%
\Gamma _{\text{cav}}\,$are necessary to achieve rapid measurement of the
system state.

\section{Classical Analogue} \label{class}

If we are to interpret the conditioned state as the best description of the
observer's knowledge of the quantum mechanical state given the results of a
series of measurements, we would expect a similarity between these equations
and classical Bayesian state observation. The analogy between the SME (\ref
{SME1}) and Kalman filtering for a classical position measurement was
discussed in \cite{milburn1996a} but only the equations for the position
probability distributions were considered. Here we formulate the continuous
time position measurement state observer for a classical harmonic oscillator
and find that there is a close analogy between the SME\ and the classical
theory for all moments of the conditioned probability distribution as long
as we restrict ourselves to Gaussian states and allow for noisy driving of 
the classical oscillator.

The problem of noisy, classical, continuous time position measurement of a
harmonic oscillator can be formulated 
\begin{mathletters}
\begin{eqnarray}
\frac{dx_{C}}{dt} &=&\omega p_{C}, \\
\frac{dp_{C}}{dt} &=&-\omega x_{C}+\sqrt{\frac{2\omega }{s}}\epsilon , \\
\frac{dQ_{C}}{dt} &=&ax_{C}+\sqrt{\frac{r}{2\omega }}\eta , \\
E[\epsilon (t)\epsilon (t^{\prime })] &=&E[\eta (t)\eta (t^{\prime
})]=\delta (t-t^{\prime }), \\
E[\epsilon (t)\eta (t^{\prime })] &=&f\delta (t-t^{\prime })
\end{eqnarray}
\end{mathletters}
We have used the same scaling of the variables as in the quantum problem. We
imagine that as well as having an imperfect measurement of the system the
oscillator is subject to a white noise force. There may be some correlation
between the oscillator (plant or process) noise and the measurement noise
and so $\epsilon $ and $\eta $ are correlated Wiener processes. As in the
quantum mechanical case the limit of small $r$ is the limit of good position
measurement. 

The continuous time theory of Kalman filtering then provides
the best estimate of the system state $\hat{x},\hat{p}$ and the second-order 
moments of
the posterior probability distribution $P(x_{C},p_{C})$ \cite{whittle1996a},
\begin{mathletters}
\begin{eqnarray}
d\hat{x}_{C} &=&\omega \hat{p}_{C}dt+a\sqrt{\frac{2\omega }{r}}V_{xx}dW, \\
d\hat{p}_{C} &=&-\omega \hat{x}_{C}dt+\left( a\sqrt{\frac{2\omega }{r}}%
V_{xp}+f\sqrt{\frac{2\omega }{s}}\right) dW, \\
\frac{1}{\omega }\frac{dV_{xx}}{dt} &=&2V_{xp}-\frac{2}{r}a^{2}V_{xx}^{2}, \\
\frac{1}{\omega }\frac{dV_{pp}}{dt} &=&-2\left( 1+\frac{2af}{\sqrt{rs}}%
\right) V_{xp}+\frac{2}{r} \nonumber \\
&&-\frac{2f^{2}}{s}-\frac{2}{r}a^{2}V_{xp}^{2}, \\
\frac{1}{\omega }\frac{dV_{xp}}{dt} &=&V_{pp}-\left( 1+\frac{2af}{\sqrt{rs}}%
\right) V_{xx}-\frac{2}{r}a^{2}V_{xx}V_{xp}.
\end{eqnarray}
\end{mathletters}
Where $dW$ is an independent Wiener increment with $dW^{2}=dt$
proportional to the
innovation process $dQ-a\hat{x}$. Note that the circumflex employed
here indicates that the quantity is an
estimate of the classical variable and not that it is a quantum operator.
The moments of the posterior probability distribution have been given the
same notation as the moments of the conditioned quantum mechanical state. 
If we make the identifications $s=r,a=\cos \phi
,f=-\sin \phi $ then this system of equations is formally identical to the
system which determines the evolution of the quantum mechanical conditioned
state Eqns (\ref{momev}). 

We see that there is 
a classical model of noisy position measurement for which the equations
of motion for the posterior probability distribution of the classical
state given by the Kalman filter 
reproduce the stochastic master equation. What is specifically quantum 
mechanical in the SME is that we cannot, even in principle,
specify the process noise and the measurement noise separately. 
Classically one could imagine isolating the system
sufficiently that $s$ is as large as we like. With $s\rightarrow
\infty$ there 
would be no momentum
disturbance on the atom and after a sufficiently long observation time the
state of the system would be determined exactly so that
$V^{\text{ss}}_{ij}=0.$ 
Clearly this
does not correspond to a quantum state. However the quantum theory of the
problem guarantees that any coupling to the system which gives position
information about the state of the system must also disturb the
momentum. This 
momentum disturbance must be sufficient that the conditioned state
always obeys
the Heisenberg uncertainty principle. Thus only some 
classically possible models of position measurement are allowed by quantum 
mechanics. For a given
level of measurement noise, the process noise must 
be at least
sufficient to ensure that the observer can never infer probability
distributions for the position and momentum which do not constitute valid
quantum states. This is a measurement-disturbance uncertainty
relation \cite{uncert} for continuous  
measurement; reducing the noise in the measurement must increase the noise
in the evolution. This backaction noise does not however behave like
classical
process noise, its properties are entirely determined by the measurement. 
If we vary the basis of the measurement on the meter --- vary $\phi $
--- then we are able to alter the correlation between the
measurement noise and the apparent process noise in the classical model.

We have found here that the symmetric moments of the conditioned state
always obey a system of equations which also describes Kalman
filtration for
a classical problem, but that altering the specific quantum measurement
alters both the observation {\em and} the process noise of the relevant
classical problem. Since the symmetric moments are the moments of the
Wigner
function, for example \cite{gardiner1991a}, the conditioned Wigner function
can be interpreted as the direct analogue of a classical posterior
probability distribution for the system. This relationship is not going to
be as straightforward for more complicated quantum systems where the conditioned 
Wigner
function can be negative and cannot be interpreted as a probability
distribution.

\section{Dependence on Initial State}  \label{mrd}

We have established above that the conditioned state of the system can in
priniciple become pure after a finite observation time. What might not be
clear is that the state which results is uniquely determined by the
measurement record. In an ideal
projective measurement the probabilities of obtaining the various
measurement results depend upon the initial state. However, once a
result has been obtained the conditional state after the measurement
depends on that result alone. In a similar way, for
continuous measurements we find that, while the initial state affects
the probability of 
obtaining particular measurement records, the conditional state
following the measurement is 
determined by the particular measurement record
which was obtained, provided that record sufficiently long.
In particular if the initial state of the system is very
poorly known to an observer then we might hope that there is
effectively a 
maximum likelihood estimate of the system state which depends only on the
measurement current and which converges to the actual system state
within $%
\tau _{\text{col}}$. If propagating the stochastic master equation with the
actual initial state of the system provides the {\em a posteriori} estimate
of the system state given the measurement results, then such a maximum
likelihood estimate would result from propagating an initial state with very
large position and momentum variances in the SME (\ref{SME1}), say $%
V_{xx}=V_{pp}=V_{0}\rightarrow \infty ,V_{xp}=0,$ which gives a nearly flat
prior probability distribution for the initial state. In this section we
demonstrate that such a strategy does indeed work. Thus the purity of the
conditioned density matrix indicates, as would be hoped, that there is only
one pure state of the system at time $t>\tau _{\text{col}}$ which is
consistent with the known sequence of measurement results.

Suppose that two observers, Alice and Bob say, postulate different
initial states of the system $\rho ^{\text{A}}(0),\rho
^{\text{B}}(0),$ which we will continue to assume to be
Gaussian. For example,
Alice may have more information than Bob about the
system in which case Bob would start  with a more
mixed initial density matrix reflecting his initial lack of
knowledge. Given that they both recieve
the same measurement record
$I(t)$ and propagate their conditioned states according to the
SME~(\ref{SME1}), it should be the case that at some time $t'$, around
$\tau _{\text{col}}$, Alice and Bob agree on the system state, so that
$\rho^{\text{A}}(t')=\rho^{\text{B}}(t')$.
From the previous section we know that after the time $\tau _{\text{col%
}}$ the second
order moments of both conditioned states will be equal
$V_{ij}^{\text{A}}=V_{ij}^{\text{B}}=V_{ij}^{\text{ss}}$ and 
so we focus on the equations for the first order moments of the
conditioned
state for each observer 
\begin{mathletters}
\begin{eqnarray}
dx^{\text{A}} &=&\omega p^{\text{A}}dt+\sqrt{\frac{2\omega }{r}}V_{xx}^{
\text{A}}dW, \\
dp^{\text{A}} &=&-\omega x^{\text{A}}dt+\sqrt{\frac{2\omega
    }{r}}V_{xp}^{
\text{A}}dW, \\
dx^{\text{B}} &=&\omega p^{\text{B}}dt+\sqrt{\frac{2\omega }{r}}V_{xx}^{
\text{B}}dW^{\text{B}}, \\
dp^{\text{B}} &=&-\omega x^{\text{B}}dt+\sqrt{\frac{2\omega
    }{r}}V_{xp}^{\text{B}}dW^{\text{B}}, \\
dQ &=&x^{\text{A}}dt+\sqrt{\frac{r}{2\omega
    }}dW=x^{\text{B}}dt+\sqrt{\frac{r}{2\omega }}dW^{\text{B}}.
\end{eqnarray}
\end{mathletters}
In this section we will omit the tildes used above to indicate that we have
scaled the position and momentum to the natural units for the harmonic
oscillator. The stochastic increment $dQ$ is the infinitesimal increment of
the measured current which both Alice and Bob have access to. We can express
the increment $dW^{\text{B}}$ in terms of the other quantities
$dW^{\text{B}%
}=dW-\sqrt{2\omega /r}(x^{\text{B}}-x^{\text{A}})dt,$ and find stochastic
differential equations for the differences between the means
$e_{x}=x^{\text{B}}-x^{\text{A}},e_{p}=p^{\text{B}}-p^{\text{A}}$, 
\begin{mathletters}
\label{errors}
\begin{eqnarray} 
de_{x} &=&\omega e_{p}dt-\frac{2\omega }{r}V_{xx}^{\text{B}}e_{x}dt \\
       & & +\sqrt{\frac{2\omega }{r}}\left
 ( V_{xx}^{\text{B}}-V_{xx}^{\text{A}}\right) dW, \\
de_{p} &=&-\omega \left( 1+\frac{2}{r}V_{xp}^{\text{B}}\right) e_{x}dt \\
       & & +\sqrt{\frac{2\omega }{r}}\left
       ( V_{xp}^{\text{B}}-V_{xp}^{\text{A}}\right) dW. 
\end{eqnarray}
\end{mathletters}
The deterministic part of this system of equations describes a damped
harmonic oscillation for $e_{x},e_{p}$ where the damping and oscillation
rates depend on the second-order moments of Bob's conditioned state. The
damping in these equations is not present in an analogous equation for $e_{x}
$ given by Mabuchi \cite{mabuchi1998a} for a free particle. The equation
adopted in \cite{mabuchi1998a} is obtained from the continuous limit of the
repeated position measurement model of Caves and Milburn \cite{caves1987a}
and does not contain a noise term in the stochastic differential equation
for $x$. In fact the omission of this term is in error and if the continuous
limit of the repeated position measurement model of Caves and Milburn is
taken correctly then the noisy contribution to $de_{x},$ which we
obtained
from the SME (\ref{SME1}), is in fact present. It is the damping which
results from this term which leads to all observers agreeing about the
conditioned state after a sufficiently long observation time. Note
that after the time $\tau _{\text{col}}\,$ Eqs~(\ref{errors}) are in
fact a system of 
ordinary differential equations since at that point the covariances of
the two conditioned states are equal.
The differences in the means then damp to the steady state values
$e_{x}=e_{p}=0\,$ indicating that Alice 
and Bob do eventually agree about the state of the system regardless
of their initial states. Thus we have shown that the conditioned state
eventually depends only on the measurement record but not the
time-scale over which this occurs.

It is
straightforward to use the It\^{o} chain rule \cite{gardiner1985a} to
find
differential equations for the expectation values of the covariance matrix
for the difference between the conditioned means of Alice and Bob 
\begin{mathletters}
\label{covev}
\begin{eqnarray} 
\frac{1}{\omega }\frac{d}{dt}\left( E[e_{x}^{2}]\right)  &=&2E[e_{x}e_{p}]-%
\frac{4}{r}V_{xx}^{\text{B}}E[e_{x}^{2}]  \nonumber \\
&&+\frac{2}{r}\left( V_{xx}^{\text{B}}-V_{xx}^{\text{A}}\right) ^{2}, \\
\frac{1}{\omega }\frac{d}{dt}\left( E[e_{p}^{2}]\right)  &=&-2E[e_{x}e_{p}]-%
\frac{4}{r}V_{xp}^{\text{B}}E[e_{x}e_{p}] \nonumber \\
&&+\frac{2}{r}\left( V_{xp}^{\text{B}}-V_{xp}^{\text{A}}\right) ^{2}, \\
\frac{1}{\omega }\frac{d}{dt}\left( E[e_{x}e_{p}]\right) 
&=&E[e_{p}^{2}]-E[e_{x}^{2}] \nonumber \\ 
&& -\frac{2}{r}V_{xp}^{\text{B}}E[e_{x}^{2}]-\frac{2}{r}V_{xx}^%
{\text{B}}E[e_{x}e_{p}] \nonumber \\
&& +\frac{2}{r}\left( V_{xx}^{\text{B}}-V_{xx}^{\text{A}}\right) %
\left( V_{xp}^{\text{B}}-V_{xp}^{\text{A}}\right) .
\end{eqnarray}
\end{mathletters}
Although we have taken the expectation value for this system of equations,
the noise terms all become zero after $\tau _{\text{col}}$ so these
ordinary
differential equations eventually describe the whole
dynamics. We are now interested in the time-scale over which the
determinant of this covariance matrix becomes zero. Unfortunately
the time dependence of the conditioned state variances prevents a closed
form solution of this system of equations and we have not found a matrix
Riccati form for the overall system. It is however straightforward to
investigate the problem numerically.

We wish to show that all observers will agree about the conditioned
state of the system in roughly $\tau_{\text{col}}$. To do this it is
sufficent to show that an arbitary observer will agree with some
preferred observer in that time. For this reason we will assume that
Alice has sufficient information to describe the state of the system
as a pure state and that she has access to sufficient earlier
measurement records that $V_{ij}^{\text{A}}(t)=V_{ij}^{\text{ss}}$.
An experimenter is going to be in Bob's position of not
having precise knowledge of or control over the preparation of the state.
We would expect that if Bob makes an accurate assessment
of his initial knowledge of the state then $E[e_{x}^{2}]$ is at first
of the 
same order as $V_{xx}$ or smaller. Since
$V_{xx}^{\text{B}}-V_{xx}^{\text{A}}$ may be large initially, the
stochastic term in the equations~(\ref{errors}) for Bob's errors
dominates for very short times and essentially 
determines a random initial condition for $e_{x}$ such that $%
E[e_{x}^{2}]\lesssim V_{xx} $. This reflects the fact that for short times
the measurement record is dominated by noise. From this point on we find
numerically that $E[e_{x}^{2}]\simeq V_{xx}$ and that all the elements
of the 
covariance matrix damp to zero within $\tau _{\text{col}}$. Thus we
have found numerically that the conditioned state
depends only 
on the measurement record after a time equal to the time over which
the conditioned state becomes pure.
 Most
importantly 
we found this to be the case even where all the variances were set to very
large initial values, the largest we tried being
$V_{xx}^{\text{B}}=V_{pp}^{\text{B}}=E[e_{x}^{2}](0)=
E[e_{p}^{2}](0)=10^{10}$.
Thus as we anticipated, 
Bob can make an accurate estimate of the state within the collapse
time even 
in the absence of accurate information about the initial state. In
this case Bob's conditioned state corresponds to the maximum
likelihood estimate discussed above. Interestingly, 
even with the pessimistic initial condition where $E[e_{x}^{2}](0)$ is
significantly larger than $V_{xx}^{\text{B}}(0)$ --- this corresponds
to Bob 
overestimating the accuracy of his estimate of the particle's position and
the difference between his estimate and the actual value being very large
--- the determinant of the covariance matrix damps on the
characteristic timescale $\tau _{\text{col}%
}$. This feature is more pronounced for larger values of $V_{xx}^{\text{B}%
}(0).$ Again this is because for large initial values of $V_{ii}^{\text{B}%
}$, Bob's estimate of the system state is essentially dependent on the
measurement current alone. These behaviours are demonstrated for the
parameters used in the previous section in Figure (\ref{fig2}).

From this we can conclude that after the time $\tau _{\text{col}}$, any
experimenter knows the state of the system regardless of how the system is
initially prepared and of how much control the experimenter has over this
process. In the next section when we introduce detection efficiency and
thermal couplings, these will just modify the second-order moments of the
conditioned state such that the steady conditioned state is no longer pure.
Since the equations for the first order moments will be unmodified all that
is necessary to reproduce the results of this section in this more general
case is to use the collapse time $\tau _{\text{col}}$ which is appropriate
for the new system. Although this will mean that the experimenter is left
with broader position and momentum probability distributions, it will not
mean that different experimenters disagree about the means of these
distributions. So regardless of the detection efficiency the conditioned
state is eventually uniquely determined by the measurement record and can
be regarded as known by the experimenter.

\section{Thermal and Detection Efficiency Effects}  \label{heat}

In order to discuss the effects of detection efficiency and other
uncontrolled coupling to a bath in the model we will add an extra momentum
diffusion term to the SME to obtain for $\phi=0$
\begin{eqnarray}
d\rho _{\text{c}}&=&-i[H_{\text{sys}},\rho _{\text{c}}]dt+2\alpha {\cal D}%
[x]\rho _{\text{c}}dt+2\beta {\cal D}[x]\rho _{\text{c}}dt \nonumber \\
&& +\sqrt{2\alpha }{\cal H}[x]\rho _{\text{c}}dW.  \label{SME2}
\end{eqnarray}
This simple modification to the master equation is intended to model several
possible imperfections in a real experiment. One contribution to $\beta $ is
the effect of detection efficiency \cite{wiseman1993a} so that $\beta $ is
at least $\alpha (1-\eta )/\eta $ where the overall detection efficiency is $%
\eta $. The effect of cavity loss through the unmonitored mirror in a cavity
QED experiment is an effective detection inefficiency. Where the loss rates
out of the two mirrors are $\kappa _{1},\kappa _{2}$, and if only the
light passing through the second mirror is detected, then we get
$\beta >\alpha \kappa _{1}/\kappa $ where $\kappa =\kappa
_{1}+\kappa _{2}$.
The best situation is if the mirror in
front of the detection apparatus has significantly higher transmission
than 
the mirror used to drive the cavity. Scattering losses of the mirror will
also be an effective detection inefficiency but these are typically much
smaller than transmission losses. In experiments with single atoms there
will also be heating due to spontaneous emission into free space which will
lead to a contribution to $\beta $ equal to $g_{0}^{2}n\Gamma /4\Delta ^{2}$
where $\Gamma $ is the free space decay rate of the excited state of the
atomic transition. Other than the restriction to the Lamb-Dicke regime this
is the largest correction to the adiabatically eliminated master equation
given above when a moderate detuning from the atomic transition is employed
as in \cite{mabuchi1998b}. The spontaneous emission contribution to the
heating is also proportional to the measurement coupling $\alpha $ such that 
$\beta _{s}=\alpha \Gamma \kappa /4g_{0}^{2}$ and to minimize the effect of
spontaneous emission we must use cavities with the largest possible value of 
$g_{0}^{2}/\kappa .$ Recall that if this rate is large the signal to noise
of the position measurement also improves. Contributions to $\beta $ in
other systems, for example the interferometric detection of the position of
a moving mirror \cite{milburn1994a}, will also come from any coupling of the
oscillator to a thermal bath. We found that the standard quantum Brownian
motion master equation \cite{gardiner1991a} led to steady conditioned states
that did not obey the Heisenberg uncertainty principle for small values of $%
r $. This a result of the non-Lindblad terms in this master equation. The
master equation we adopt here solves this problem by considering only
coupling to very high temperature thermal baths for which the thermal
contribution to $\beta $ is$\frac{\gamma k_{B}T}{\hbar ^{2}},$ where $\gamma 
$ is the coupling rate to the thermal reservoir and $T$ is the temperature
of the bath. This should be an adequate description of heating in the
experiment as long as the bath to which the system is coupled is of
sufficiently high temperature that only the diffusive evolution is
significant for the timescales of interest.

There is now no SSE equivalent to this SME (\ref{SME2}) and pure states
become mixed during the evolution. However the SME\ continues to preserve
Gaussian states so the previous calculation for the evolution of state can
be straightforwardly modified. The second-order moments of the conditioned
state still reach a steady state and we can easily find an expression for
the steady conditioned state phase space area $A\,$is given by 
\begin{mathletters}
\begin{eqnarray}
A^{\text{ss}} &=&\sqrt{1+\beta /\alpha }=q\geq 1/\sqrt{\eta }, \\
s(\rho _{\text{c}}^{\text{ss}}) &=&1-1/q\geq 1-\sqrt{\eta }, \\
S(\rho _{\text{c}}^{\text{ss}}) &=&\frac{q+1}{2}\ln (q+1)-\frac{q-1}{2}\ln
\left( q-1\right) -\ln 2.
\end{eqnarray}
\end{mathletters}
The linear and von Neumann entropies of the steady conditioned states are
plotted for a range of detection efficiencies in Fig.\ (\ref{fig3}). Even
though we have effectively coupled the oscillator to an infinite temperature
bath the conditioned steady states have in some sense a finite temperature
but stochastically varying mean values of position and momentum.

We will consider some limiting cases here for the steady state variances in
this more general situation. Where the measurement is strong ($r\ll 1$) the
position variance is insensitive to these imperfections, $V_{xx}^{\text{ss}%
}\simeq \sqrt{r}\sqrt{q},V_{xp}^{\text{ss}}\simeq q,V_{pp}^{\text{ss}}\simeq
2\sqrt{q}^{3}/\sqrt{r}$. If the dynamics dominate ($r\gg 1$) then the
position and momentum variances have the same dependence on $\beta $ as $A$, 
$V_{xx}^{\text{ss}}\simeq q,V_{xp}^{\text{ss}}\simeq q^{2}/r,V_{pp}^{\text{ss%
}}\simeq q$.

Finally the whole time evolution of the second-order moments of the
conditioned state can be determined by solving the matrix Riccati equation.
Again we will just consider the position variance as a function of time
where $V_{xx}(0)=V_{pp}(0)=V_{0}>1,V_{xp}(0)=0$
\begin{mathletters}
\begin{eqnarray}
V_{xx}(t) &=&\frac{%
\begin{array}{c}
q^{2}V_{0}\left( c^{2}\cosh 2bt+b^{2}\cos 2ct\right)  \\ 
+ q\left( V_{0}^{2}+q^{2}\right) \left( c\sinh 2bt-b\sin 2ct\right)
\end{array}
}{%
\begin{array}{c}
q^{2}\left( b^{2}+c^{2}\right) +qV_{0}\left( c^{3}\sinh 2bt+b^{3}\sin
2ct\right)  \\ 
+ \left( V_{0}^{2}+q^{2}\right) \left( c^{2}\sinh ^{2}bt-b^{2}\sin
^{2}ct\right)
\end{array}
}, \\
b &=&\frac{1}{\sqrt{2}}\sqrt{\sqrt{1+\frac{4q^{2}}{r^{2}}}-1}, \\
c &=&\frac{1}{\sqrt{2}}\sqrt{\sqrt{1+\frac{4q^{2}}{r^{2}}}+1}=q/V_{xx}^{%
\text{ss}}
\end{eqnarray}
\end{mathletters}
and we have scaled time by the harmonic oscillation frequency $\omega $. As
before when $2bt\gg 1\,$the system reaches its steady state value and the
non-linearity of the terms describing the conditioning of the system state
mean that the time for this to occur is independent of the initial state.
Interestingly this time is in fact shorter than was required to purify the
conditioned state in the case of ideal detection. This is essentially
because the extra noise means that past observations become irrelevant more
quickly, not leaving enough time to determine a pure state completely.While
the steady state is reached increasingly fast it corresponds to an
increasingly high effective temperature. When $r>1$\thinspace $\,$this time
to reach the steady state is $\tau _{\text{s}}\simeq 2r/q\omega =m\omega
/q\hbar \alpha =\tau _{\text{col}}/q$ since in this regime $b\simeq q/r\,$.
As was noted in the previous section the time for the conditioned state
variances to reach their steady state is also the time that is necessary for
different observers to agree about the system state. The time evolution of
some of these quantities is plotted in Figure (\ref{fig4}).

\section{Conclusions} \label{discuss}

In this paper we have established that for a simple class of systems quantum
trajectory theories allow the determination of a unique post-measurement
state that depends only on the measurement results over a finite time. We
have discussed the effects of experimental imperfections on this state
determination and the analogy to classical state observation. 
The systems
to which this analysis is applicable have the property that the conditioned
density matrix is at all times Gaussian and its evolution is exactly
that of 
the posterior probability distribution for an appropriate classical state
observer. While we have considered only the case of position
measurement the same treatment will be applicable to these other linear
systems. Clearly in more complicated systems, such as the resonant
interaction of an atom with the single mode of an optical cavity, this will
not always be the case and the conditioned Wigner function will
sometimes be non-positive. We would however expect that the central
result 
we have shown here, the purity of the conditioned state after
sufficiently 
long continuous observation and the dependence of this state on the
initial 
state only through the measurement results, will still hold for these
more 
interesting and complicated systems. This will however require numerical
simulation of the stochastic master equation for such systems. Another
feature that should 
generalize is the interpretation of the SME as
a state observer --- presumably an optimal one --- for the quantum
system.

\section*{Acknowledgements}

A. D. would like to thank Kurt Jacobs for continuing stimulating
discussions 
on continuous measurement theory and in particular for suggesting that it
would be interesting to calculate the purity of conditioned states
with and 
without thermal couplings. It is also a pleasure to acknowledge
discussions 
with Hideo Mabuchi. This work is supported by the Marsden Fund of the
Royal 
Society of New Zealand.

\begin{figure}[tbp]
\caption{The time evolution of (a) the unitless second-order moments $%
V_{xx},V_{xp},V_{pp}$ and (b) the two entropies of the conditioned state of
the harmonic oscillator under continuous position measurement as described
in the text. Note the different time axes for the entropies. Time is
measured in units of the harmonic oscillator angular frequency while $r=20$.
The initial state is $V_{xx}=V_{pp}=20,V_{xp}=0$ corresponding to a thermal
state of the oscillator.}
\label{fig1}
\end{figure}

\begin{figure}[tbp]
\caption{The time evolution of Bob's mean-squared error in position $E[e_{x}^{2}]$,
Bob's conditioned state variance $V_{xx}^{\text{B}}$ is also plotted for
comparison. In (a) the initial values are $E[e_{x}^{2}]=E[e_{p}^{2}]=5,E[e_{x}e_{p}]=0,%
V_{xx}^{\text{B}}=V_{pp}^{\text{B}}=200,V_{xp}^{\text{B}}=0$ and 
as discussed in the text the mean squared error, although originally small,
rapidly becomes of the same order as the conditioned state variance. In (b)
the initial
values are $E[e_{x}^{2}]=E[e_{p}^{2}]=10^{10},E[e_{x}e_{p}]=0,V_{xx}^{\text{B%
}}=V_{pp}^{\text{B}}=10^{10},V_{xp}^{\text{B}}=0\,$ and over a time roughly
equal to $\tau _{\text{col}}$ the mean squared error approaches zero and the
conditioned position variance approaches its steady state. Note the
different time-scales in the two graphs.}
\label{fig2}
\end{figure}

\begin{figure}[tbp]
\caption{The linear and von Neumann entropies of the steady conditioned state are
plotted against effective detection efficiency. Extraneous heating of the
oscillator is also described by $\eta$ as discussed in the text.}
\label{fig3}
\end{figure}

\begin{figure}[tbp]
\caption{(a) The time evolution of the unitless second-order moments $%
V_{xx},V_{xp},V_{pp}$ under continuous imperfect position measurement as
described in the text. Time is measured in units of the harmonic oscillator
angular frequency while $r=20$ and $q=5$. The initial state is $%
V_{xx}=V_{pp}=20,V_{xp}=0$ corresponding to a thermal state of the
oscillator and for comparison the evolution of the conditioned position
variance is plotted for the case of perfect detection. As discussed in
the text the conditioned second-order moments reach steady state more
rapidly when the detection is imperfect. (b) The mean-square error in Bob's
estimate of the conditioned state mean position and Bob's conditioned state
position variance for imperfect continuous position measurement $r=20,q=5.$
The initial values are $%
E[e_{x}^{2}]=E[e_{p}^{2}]=10^{10},E[e_{x}e_{p}]=0,V_{xx}^{\text{B}}=V_{pp}^{%
\text{B}}=10^{10},V_{xp}^{\text{B}}=0$. Even where the observation is
imperfect the conditioned state is eventually independent of the initial
state of the system.}
\label{fig4}
\end{figure}
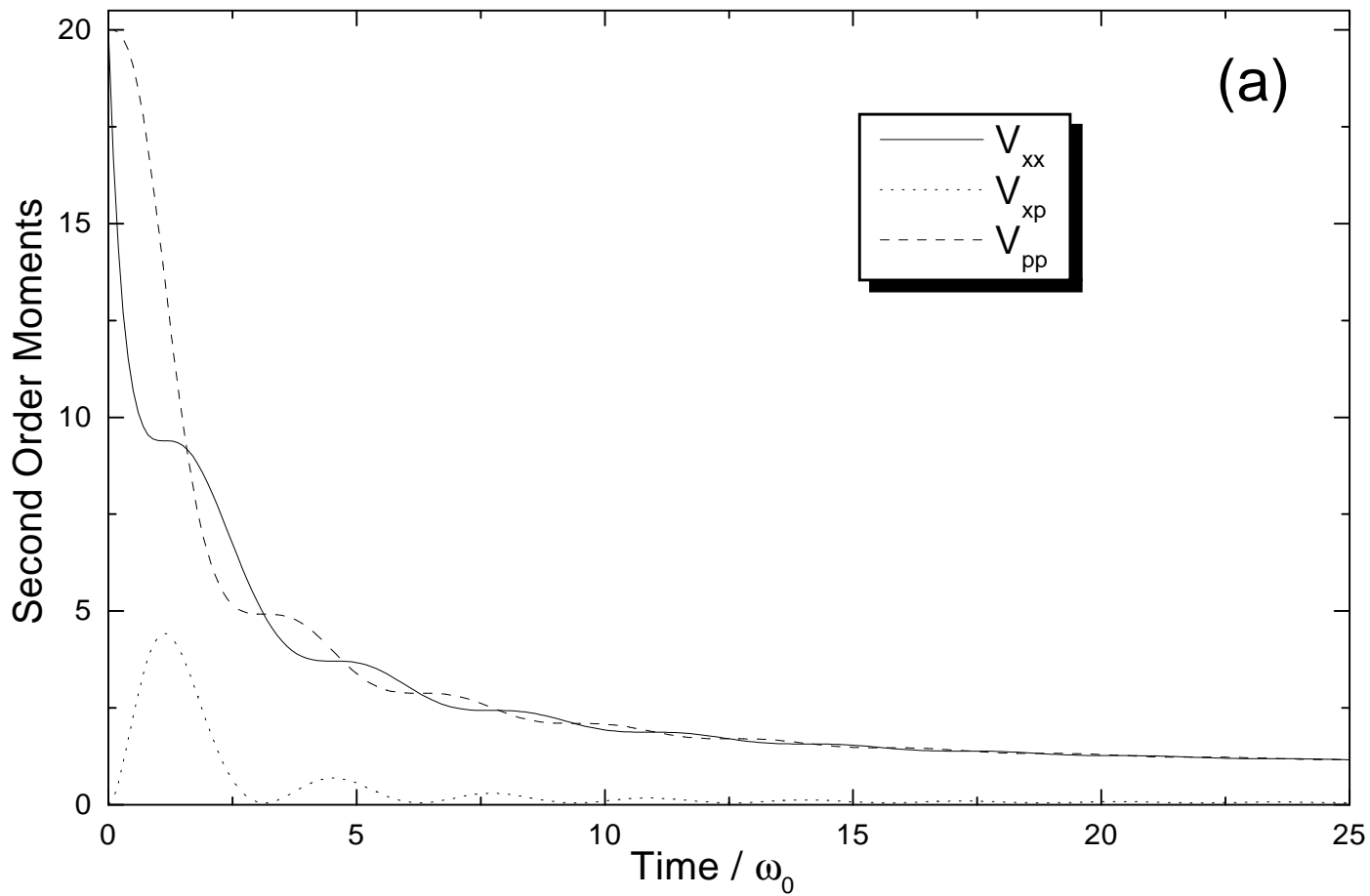
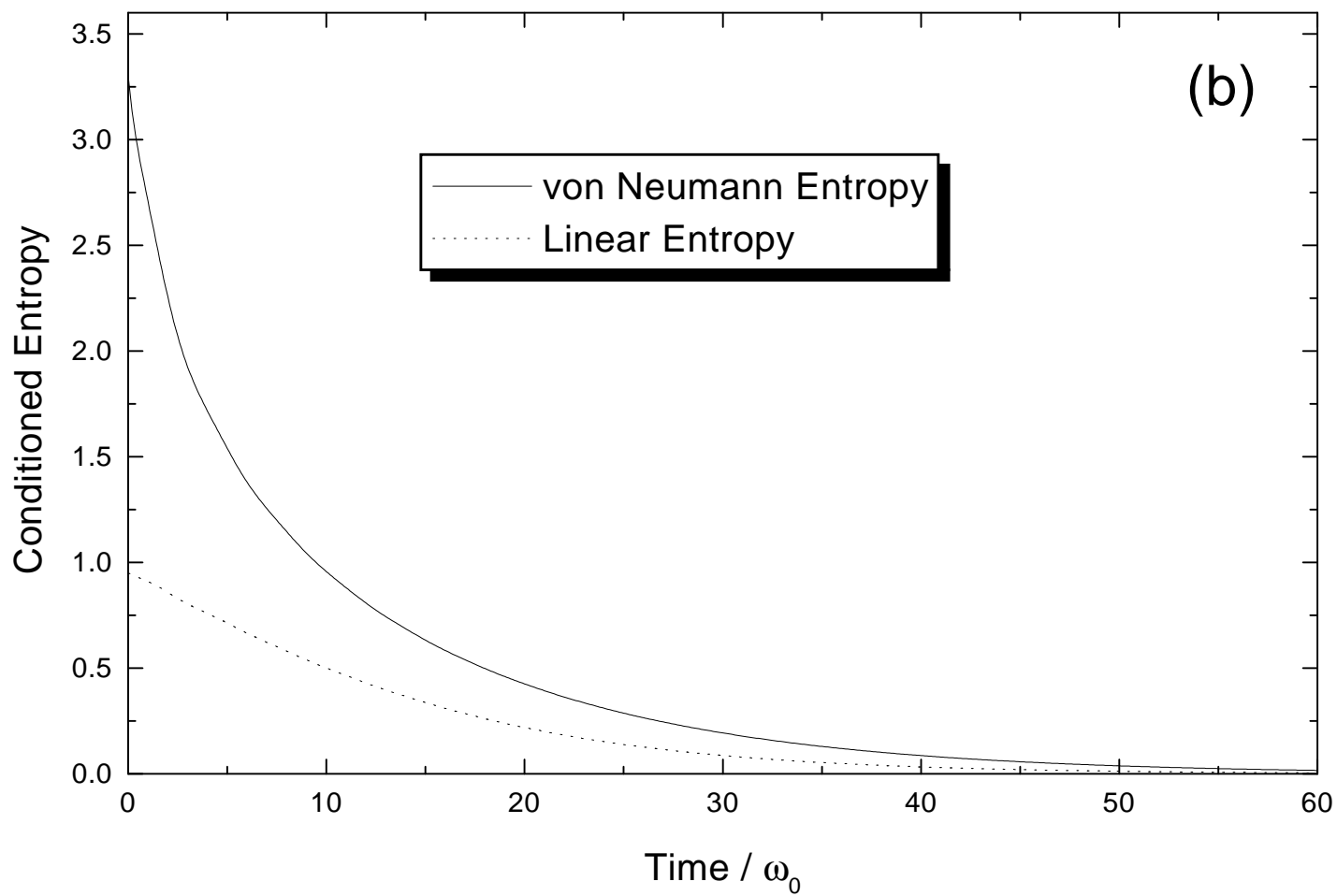
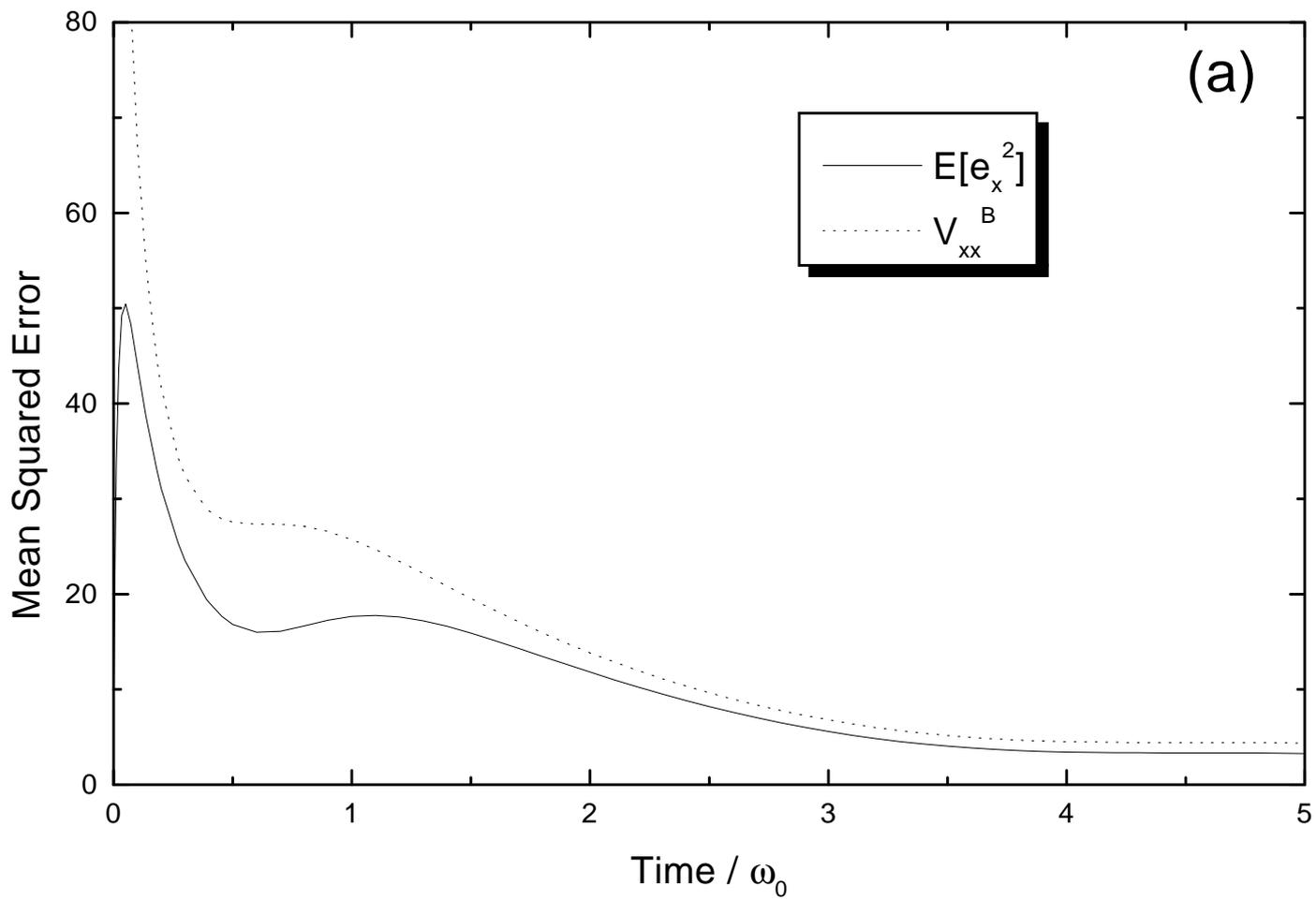
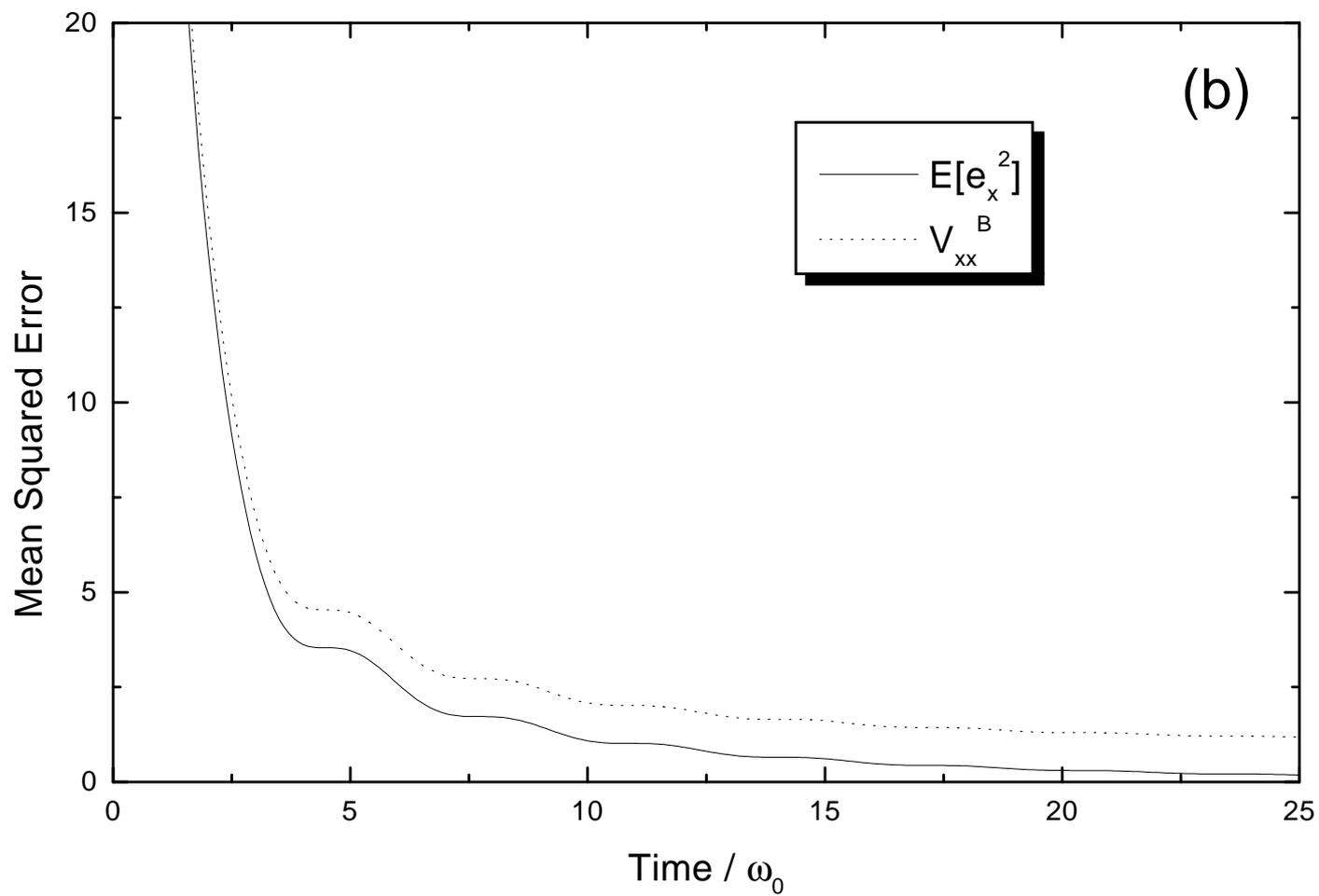
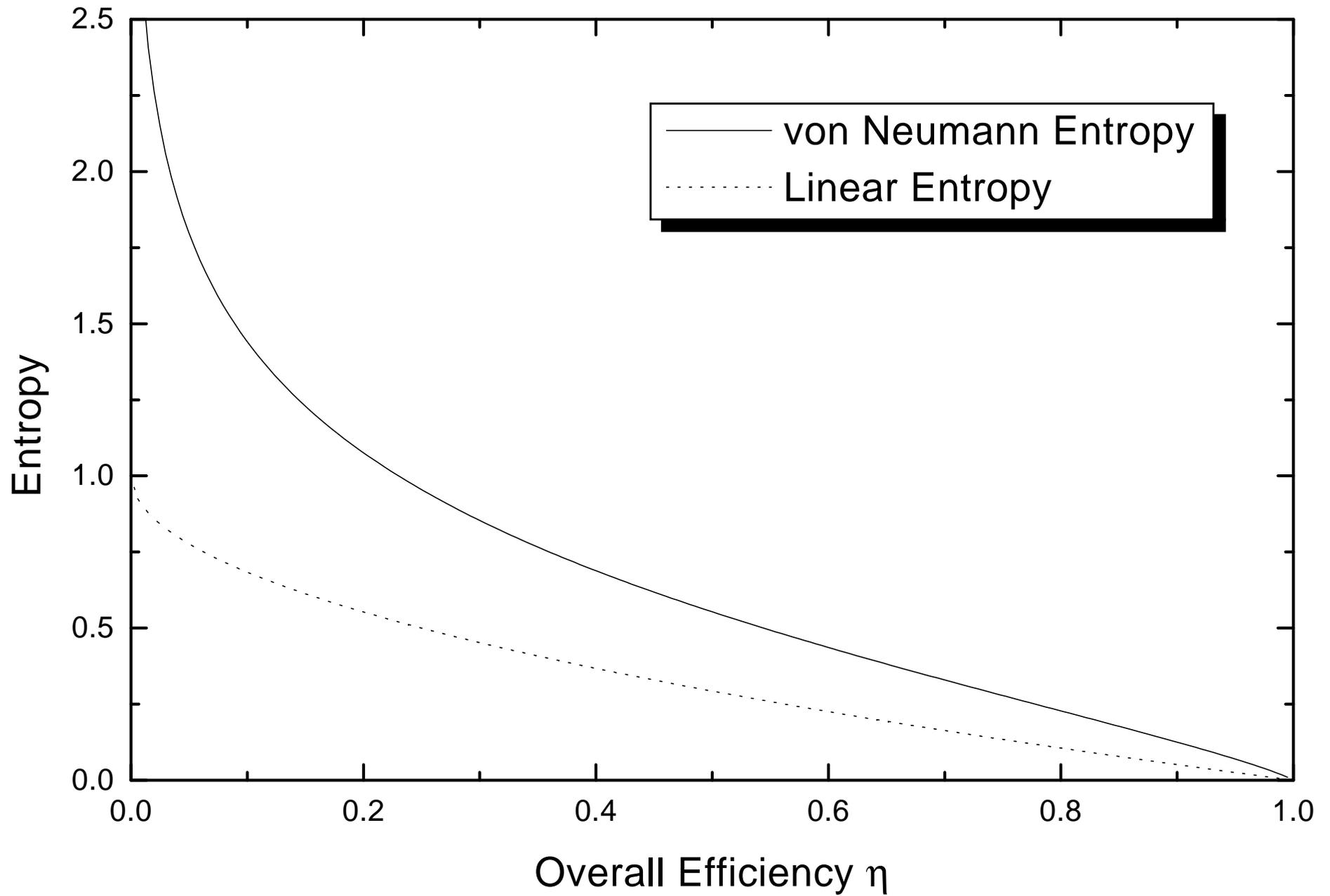
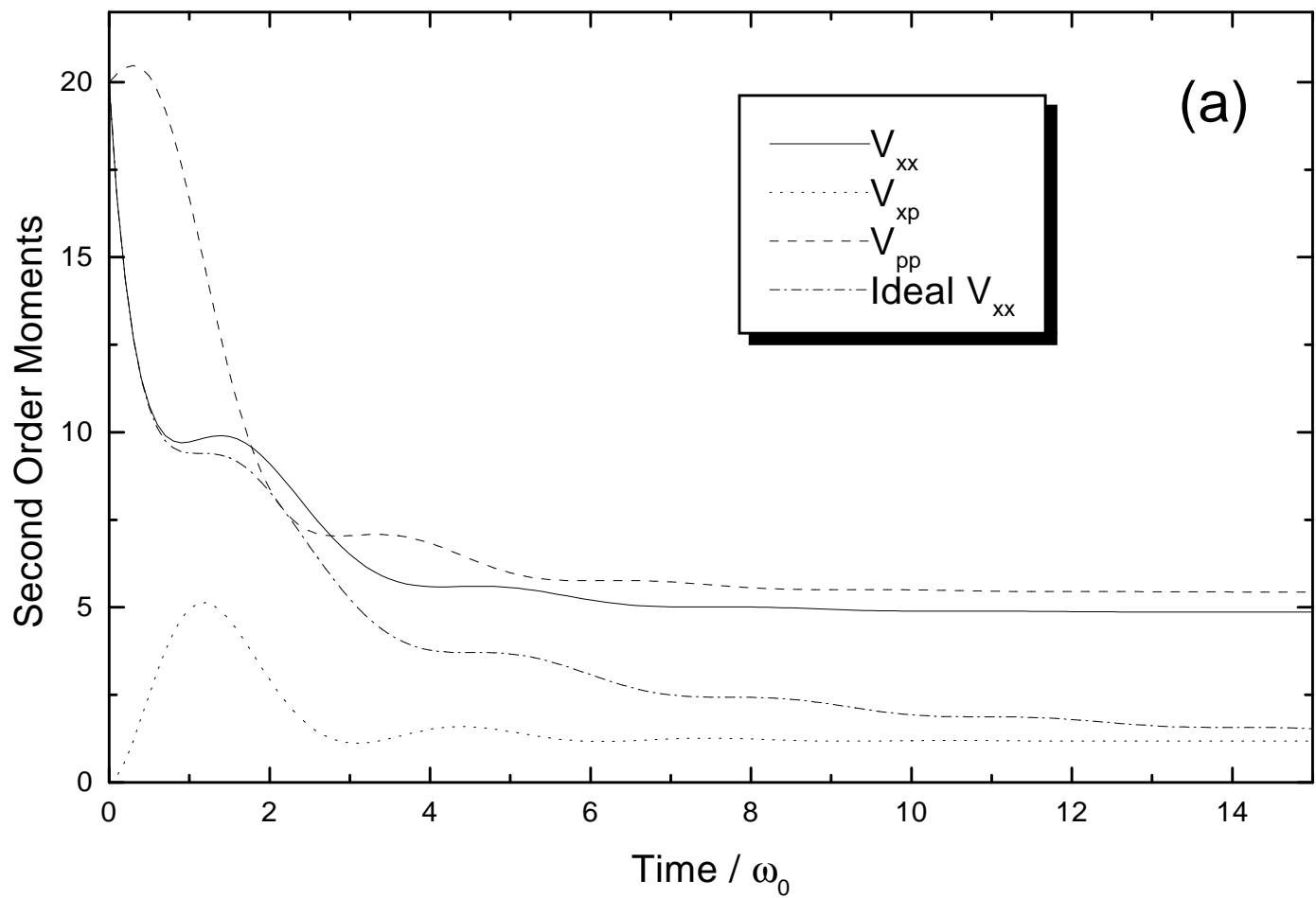
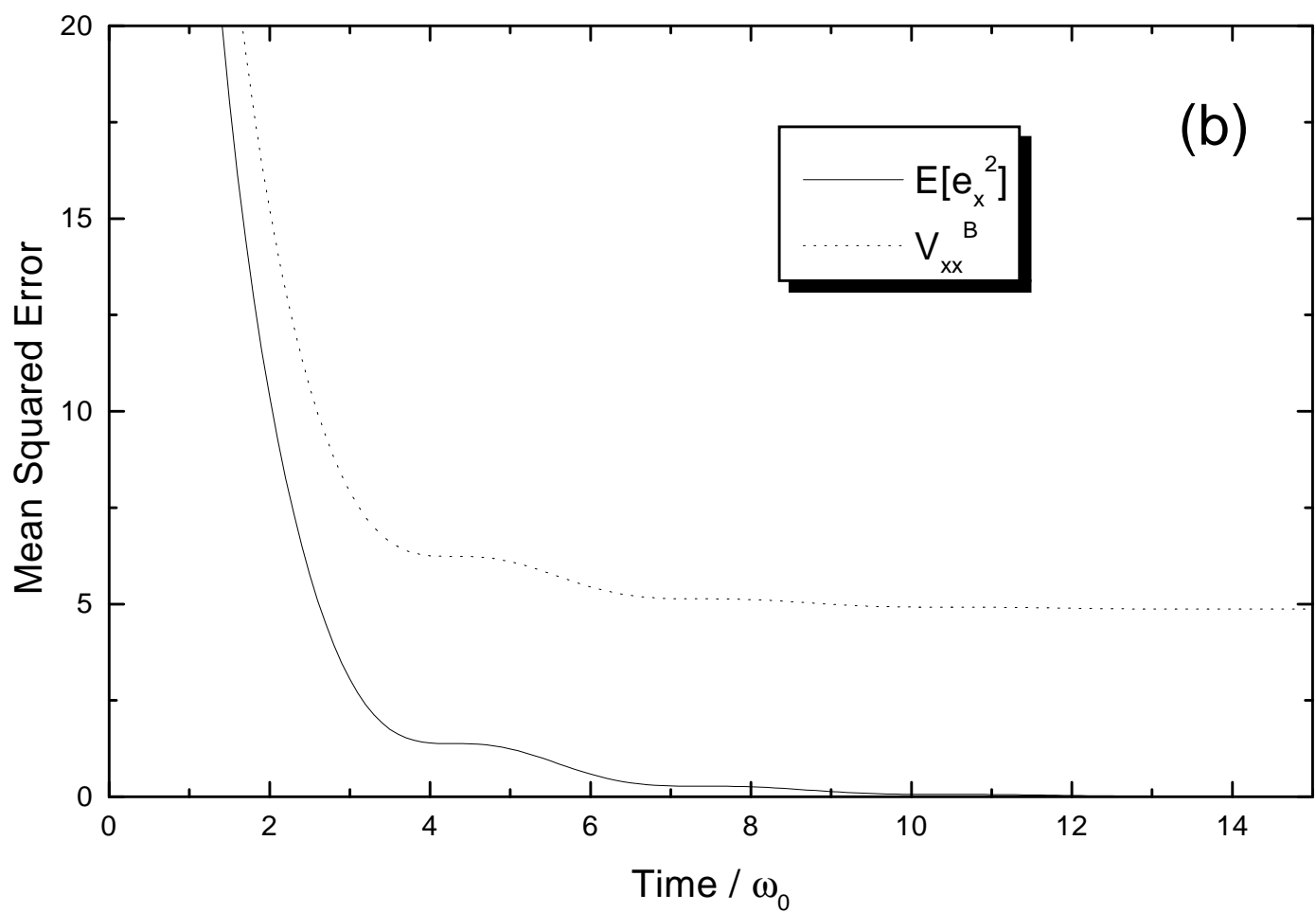
\end{document}